\documentclass[prl, twocolumn, superscriptaddress, nofootinbib, amsfont, graphicx, preprintnumbers, aps]{revtex4-2}

\UseRawInputEncoding
\usepackage{color, graphicx, epsfig}
\usepackage{ifpdf}
\usepackage{amsmath}
\usepackage{booktabs}
\usepackage{bm}
\usepackage[english]{babel}
\usepackage{amssymb}
\usepackage{braket}
\usepackage{hyperref}
\usepackage{enumerate}
\usepackage{url}
\usepackage{multirow}
\usepackage{threeparttable}
\usepackage{makecell}
\usepackage{setspace}
\usepackage{float}

\usepackage{soul}

\usepackage{slashed}

\usepackage{changes}

\begin{document}
\preprint{CPTNP-2025-033}

\title{Enhanced Cosmic-Ray Cooling in AGN from Dark Matter Deep Inelastic Scattering}

\author{Linjie Li}
\email{linjieli@njnu.edu.cn}
\affiliation{Department of Physics and Institute of Theoretical Physics, Nanjing Normal University, Nanjing, 210023, China}

\author{Chih-Ting Lu}
\email{ctlu@njnu.edu.cn}
\affiliation{Department of Physics and Institute of Theoretical Physics, Nanjing Normal University, Nanjing, 210023, China}
\affiliation{Nanjing Key Laboratory of Particle Physics and Astrophysics, Nanjing, 210023, China}

\author{Arvind Kumar Mishra}
\email{arvindm215@gmail.com}
\affiliation{Department of Physics and Institute of Theoretical Physics, Nanjing Normal University, Nanjing, 210023, China}
\affiliation{Nanjing Key Laboratory of Particle Physics and Astrophysics, Nanjing, 210023, China}

\author{Liangliang Su}
\email{liangliang.su@kit.edu}
\affiliation{Institute for Astroparticle Physics, Karlsruhe Institute of Technology, D-76131 Karlsruhe, Germany}

\author{Lei Wu}
\email{leiwu@njnu.edu.cn}
\affiliation{Department of Physics and Institute of Theoretical Physics, Nanjing Normal University, Nanjing, 210023, China}
\affiliation{Nanjing Key Laboratory of Particle Physics and Astrophysics, Nanjing, 210023, China}

             
\begin{abstract}

The diffusion of high-energy cosmic rays (CRs) through the dark matter (DM) spikes of active galactic nuclei entails significant energy loss via interactions with DM. While previous studies of sub-GeV DM have focused on elastic scattering, this process becomes insufficient at higher proton energies and DM masses. In this work, we investigate the CR-DM deep inelastic scattering (DIS) as mediated by a vector portal. We calculate the DIS contribution to the CR energy loss rate and derive stringent exclusion limits on the CR-DM scattering cross-section for DM masses between $10^{-6}$ GeV and $1$ GeV. For higher CR energies and mediator masses, the resulting CR cooling timescales are reduced by orders of magnitude after involving the DIS contribution, producing stringent constraints that surpass most of current experimental limits.

\end{abstract}

\maketitle

\textbf{\textit{Introduction:}}
Cosmological observations robustly establish that dark matter (DM) constitutes a dominant component of the universe's energy budget~\cite{Bertone:2004pz, ParticleDataGroup:2024cfk}, and its particle identity still persists as a fundamental enigma. Conventional direct detection experiments constrain well-motivated DM candidates including Weakly Interacting Massive Particles (WIMPs) through nuclear recoils~\cite{Lee:1977ua, Jungman:1995df}, however, due to kinematic thresholds, their sensitivity diminishes drastically below the sub-GeV scale of DM mass~\cite{Essig:2011nj, Essig:2012yx, Aprile:2012zx, Kouvaris:2016afs, PandaX-II:2017hlx, Dolan:2017xbu, Su:2020zny, Flambaum:2020xxo, Wang:2021jic, Bell:2021ihi, Wang:2021oha, Elor:2021swj, Li:2022acp, Shakeri:2022dwg, PandaX:2023tfq, DAMIC-M:2023gxo, Gu:2023pfg, SENSEI:2023zdf, DeMarchi:2024riu, Guo:2024sqh, Balan:2024cmq, Lin:2025pqh}.

For sub-GeV DM candidates, a variety of production and evolution mechanisms exist beyond the simple $2\to 2$ annihilation of the WIMP paradigm, e.g. number-changing $3\to 2$ processes in the dark sector~\cite{Hochberg:2014dra,Fitzpatrick:2020vba,Ho:2022erb}, asymmetric DM~\cite{Bhattacherjee:2013jca,Izaguirre:2015yja,Ho:2022tbw}, and freeze-in production~\cite{Frangipane:2021rtf,Bhattiprolu:2022sdd}. In these scenarios the present-day annihilation rate is typically negligible, so direct detection and accelerator experiments may play a central role in probing light DM~\cite{Izaguirre:2015yja,Knapen:2017xzo}. However, the sensitivity of these experiments is limited by detector size and technical issues \cite{MarrodanUndagoitia:2015veg, Cebrian:2022brv}. It is therefore necessary to investigate other extreme environments which may offer enhanced sensitivity to the non-annihilating nature of DM.

Active galactic nuclei (AGNs) are considered optimal laboratories for cosmic ray (CR) and DM interactions to address this challenge in the sub-GeV DM regime~\cite{Blandford:1979za, Berezinsky:1996wx, Peirani:2016qvp, IceCube:2018cha, Russell:2019snh, Niblaeus:2019gjk, 2021PhRvD.103l3018Z, Mbarek:2022nat, Herrera:2023nww, Gustafson:2025dff}. Multi-messenger emissions (${\gamma}$-rays/neutrinos) and direct neutrino penetration enable direct probes of central-engine processes~\cite{Inoue:2011bp, Tamborra:2014xia, Fiorillo:2023dts}.The powerful techniques developed in these multi-messenger studies have concurrently established AGNs and other astrophysical environments as premier laboratories for DM searching~\cite{Essey:2009ju, Bernal:2012qh, e-ASTROGAM:2017pxr, Bringmann:2018cvk, Buonocore:2019esg, DAMPE:2019gys, Hinton:2019etp, IceCube-Gen2:2020qha, SuperCDMS:2022kse, Lin:2022dbl, Ruszkowski:2023rzd, Ehlert:2024lji, IceCube:2024ayt, Blanco-Mas:2024ale, CDEX:2024rfi, Wang:2025uwh}. The adiabatic growth of supermassive black hole (SMBH) can gravitationally compress the surrounding DM to form a DM spike, thereby enhancing the DM density in its vicinity~\cite{Gondolo:1999ef, Quinlan:1994ed, Ullio:2001fb}. As the black hole slowly accumulates mass, the surrounding dark matter particles experience a gravitational potential that evolves gradually, preserving their orbital actions. This process generates a dense DM spike, which serves as a natural ideal place for the DM-baryon scattering. In such a dense spike environment, the propagation of CRs through the DM spike results in energy loss to DM particles and consequent CR cooling~\cite{Cappiello:2018hsu}. This mechanism has been used to explore the constraints for DM models with a scalar mediator~\cite{Herrera:2023nww}, inelastic DM scenarios~\cite{Gustafson:2024aom}, and gauged $U(1)$ extended DM models~\cite{Mishra:2025juk}. Due to a single scattering between CR and DM, CR cooling constraints are robust in compare to those generated by CR-boosted DM which requires multiple interactions~\cite{Pinzke:2011ek, Wang:2019jtk, Leane:2020wob, PandaX-II:2021kai, Xia:2021vbz, Wang:2021nbf, CDEX:2022fig, Maity:2022exk, Liang:2024xcx, LZ:2025iaw, Sun:2025gyj}.

Notably, the CR cooling efficiency intrinsically depends on energy and momentum transfer during DM-CR scattering~\cite{Abdallah:2015ter, Boveia:2018yeb}. When protons in CR transfer significant momentum to DM, $Q^2 \gtrsim 1~\mathrm{GeV}^2$ (where $Q^{2}=- (p-p^{\prime})^2$, in which $p$ and $p^{\prime}$ are the four-momenta of the initial and final protons, respectively), approximations ignoring proton internal structure and inelastic scattering channels become inadequate. Previous treatments of DM-proton cross sections primarily assumed elastic scattering with simplified dipole form factors~\cite{Hohler:1976ax, Engel:2013lsa, Planck:2015mvg, Anzuini:2021lnv}, potentially overestimating CR energy loss rates by neglecting inelastic scattering contributions. In recent years, the importance of inelastic scattering processes has been recognized across various dark matter studies. When the momentum transfer becomes comparable to or exceeds the characteristic scale of the target’s internal structure, inelastic channels can no longer be neglected. For instance, in direct detection experiments, boosted dark matter scenarios can naturally dominate over elastic scattering due to kinematic enhancements or the suppression of elastic channels~\cite{Bringmann:2018cvk, Nagata:2014aoa, Filimonova:2022pkj, Su:2023zgr}. Similarly, in neutron stars, the dominant mode of interaction with high kinetic energy DM is inelastic scattering~\cite{Su:2024flx}.

Motivated by these precedents, we turn to the AGN environment, where TeV-PeV CR protons in sources like NGC~1068 can undergo scattering with momentum transfer reaching $Q^2 \gtrsim 1~\mathrm{GeV}^2$. In this high-$Q^{2}$ regime, the scattering process is dominated by DIS off the proton's constituent quarks, leading to significantly different energy transfer rates compared to the elastic scattering approximation. Consequently, previous treatments restricted to low momentum transfer are insufficient for robustly constraining high energy CR-DM scattering~\cite{Mishra:2025juk}. 


In this work, we explore the constraints on DM properties involving both elastic and inelastic scattering processes for CR protons and DM using the CR cooling criterion in DM spike. We assume a minimal fermionic DM model in which DM couples to standard model (SM) particles through vector portal interactions. This theoretical framework self-consistently quantifies the transition from elastic scattering dominance at low momentum transfer to inelastic scattering regimes at high momentum transfer, representing a pivotal advancement for modelling CR cooling in AGN jets. Our findings reveal that, for heavier mediators, the influence of inelastic scattering on the DM energy loss rate closely approaches and even surpasses that of elastic scattering, particularly in parameter space characterized by heavier DM mass.

This paper is structured as follows. Section~II describes the DM spike and the CR cooling formalism. Section~III introduces the vector portal model and presents the elastic and DIS cross sections separately. In Section~IV, the CR cooling timescales and exclusion limits on the DM–proton cross section are derived. Section~V concludes with a summary and outlook.

\textbf{\textit{CR Cooling Effect via CR-DM Interactions:}} 
The SMBH at the center of AGN can accelerate CR protons to high energies; these protons then produce high-energy neutrinos and photons via \(pp\) and \(p\gamma\) interactions~\cite{Murase:2022dog}. Multimessenger observations of the NGC 1068 allow us to estimate the energy-dependent cooling time of CR protons primarily through standard model processes, including \(pp\), \(p\gamma\), and Bethe-Heitler pair production~\cite{Murase:2019vdl, Fiorillo:2024akm, Fiorillo:2025ehn}. For NGC~1068, whose host galaxy appears to have a elatively quiet merger history~\cite{Merritt:2006mt, 2021ApJS..257...61Y}, a dense DM spike is expected to survive to the present day. The spike profile follows a power-law form from adiabatic compression, parameterized as $\rho_\chi(r) \propto r^{-\gamma_{\text{sp}}}$ within a characteristic spike radius $R_{\text{sp}}$. The spike slope $\gamma_{\text{sp}}$ is given by $\gamma_{\text{sp}} = (9-2\gamma)/(4-\gamma)$~\cite{Meighen-Berger:2025hrq, Gondolo:1999ef}. Here, $\gamma$ is the inner slope of the initial Navarro-Frenk-White (NFW) profile~\cite{Navarro:1996gj}, which typically lies in the range $0.5 \lesssim \gamma \lesssim 1.5$. For the canonical NFW case with $\gamma = 1$, one obtains $\gamma_{\text{sp}} = 7/3$. In this work, we focus on the NGC~1068 environment, and adopt a benchmark central density of $\rho _{\chi }=5 \times 10^{18}~\mathrm{GeV}/\mathrm{cm}^{3}$ based on the dedicated analysis of NGC 1068-like systems in Ref.~\cite{Herrera:2023nww}.

In addition to the SMBH's parameters, the DM spike profile also depends on DM microphysics; for non-annihilating DM particles, DM spike density is high, whereas it decreases for annihilating or self-interacting DM scenarios~\cite{Chiang:2019zjj, Alvarez:2020fyo}. Therefore, the enhanced DM density near the SMBH's vicinity can influence the CR propagation. CRs passing through the DM spike will scatter off DM particles and transfer their energy to them. In case of large energy transfer during CR-DM scattering in DM spike, the cooling timescale via CR proton-DM scattering can be smaller than the SM prediction, therefore it strongly modify the cooling timescale, and spectrum of CRs.

The cooling timescale for CRs undergoing deep inelastic scattering with DM particles follows an analogous formulation to the elastic scattering case. Building upon the standard definition of energy loss rate~\cite{Ambrosone:2022mvk}, we extend this framework to encompass inelastic scattering processes~\cite{Gustafson:2024aom}:
\begin{equation}
    t_{\chi p}=\left[-\frac{1}{E}\left(\frac{\mathrm{~d} E}{\mathrm{~d} t}\right)_{\chi p}\right]^{-1},
\label{eq:time}
\end{equation}
where $E$ is the energy of CRs and $\dot{E}_{\chi p} \equiv \left(\frac{\mathrm{~d} E}{\mathrm{~d} t}\right)_{\chi p}$ denotes the CR-DM collision energy loss rate. This rate is defined as
\begin{equation}
    \dot{E}_{\chi p}=-\frac{\rho_{\chi}}{m_{\chi}} \int_{0}^{T_{\chi}^{\max }} \mathrm{d} T_{\chi} T_{\chi} \frac{\mathrm{d} \sigma}{\mathrm{~d} T_{\chi}},
    \label{eq:energy loss}
\end{equation}
with $m_{\chi}$ the DM mass, $\rho_{\chi}$ the average density of DM particles in the CR production region, and ${\mathrm{d} \sigma}/{\mathrm{~d} T_{\chi}}$ the differential DM-proton cross section for final-state DM kinetic energy $T_{\chi}$. The kinematics of energy transfer differ fundamentally between elastic and DIS processes. In the context of elastic scattering, the maximum kinetic energy imparted to the DM particle is derived using the four-momentum conservation as
\begin{equation}
    T_{\chi,el}^{\max }=\frac{2 T^{2}+4 m_{p} T}{m_{\chi}}\left[\left(1+\frac{m_{p}}{m_{\chi}}\right)^{2}+\frac{2 T}{m_{\chi}}\right]^{-1}, 
\end{equation}
where $T=E-m_{p}$ is the proton kinetic energy. For DIS in CR-DM scattering, the final state contains a multi-hadron system in addition to the DM particle. In the rest frame of the initial DM particle, energy conservation gives $T_\chi\leq T+m_p-W\leq T$, with a multi-hadron system of invariant mass $W\ge m_p$. The DIS contribution to Eq.~\ref{eq:energy loss} is therefore evaluated with $T$ as the outer integration limit and the inclusive differential cross section restricted to the allowed phase space.

\textbf{\textit{CR-DM Elastic and Deep Inelastic Scattering:}} 
The functional form of ${\mathrm{d} \sigma}/{\mathrm{~d} T_{\chi}}$, whether elastic or deep inelastic scattering, is determined by the underlying particle physics model. For simplicity, we assume DM to be a Dirac fermion $\left(\chi \right)$, interacting with a proton via the mediator dark photon $\left(A_{\mu }'\right)$. The corresponding Lagrangian is given by
\begin{equation}
    \mathcal{L}=g_{\chi} \bar{\chi} \gamma^{\mu} \chi A_{\mu}^{\prime}+\sum_{q} g_{q} \bar{q} \gamma^{\mu} q A_{\mu}^{\prime}~,
\label{eq:Lagrangian}
\end{equation}
where $q$ for quarks, with coupling constants $g_{\chi}$ (DM-dark photon) and $g_{q} = Q_{q} e \epsilon$ (quarks-dark photon), where $Q_{q}$ depends on the electrical charge of quarks and $\epsilon$ the kinetic mixing parameter of new $U(1)$ and the SM $U(1)_Y$ gauge symmetries~\cite{Holdom:1985ag, Essig:2013lka}.

In the DM-proton elastic scattering, the proton can be treated as point-like particle in the regime of low transfer momentum, $q^2\to 0$, and effects of its internal structure are negligible. However, as the momentum transfer increases, the internal structure of the nucleon becomes progressively relevant, for instance through its charge and magnetization distributions under the Lagrangian of Eq.~(\ref{eq:Lagrangian}). In this region, the effective DM-proton interaction Lagrangian can be defined by the electromagnetic form factors $F_1(Q^2)$ and $F_2(Q^2)$~\cite{Perdrisat:2006hj}, 
\begin{equation}
  \bra{N(p^{\prime})} \sum_{q} g_{q} \bar{q} \gamma^{\mu} q A_{\mu}^{\prime} \ket{N(p)} = g_p A_{\mu}^{\prime} \bar{N} \Gamma^{\mu} N,
  \label{eq:effint}
\end{equation}
\begin{equation}
\mathrm{with} \quad   \Gamma^{\mu}=\gamma^{\mu} F_{1}\left(Q^{2}\right)+\frac{i}{2 m_{p}} \sigma^{\mu \nu} q_{\nu} F_{2}\left(Q^{2}\right),
\end{equation}
where $g_p = \epsilon e $ denotes the coupling between the dark photon and the proton. The squared momentum transfer is given by $Q^{2}=-q^{2} =- (p-p^{\prime})^2$, where $p$ and $p^{\prime}$ are the four-momenta of the initial and final protons, respectively. At low $Q^{2}$, the Pauli form factor $F_{2}$ is typically suppressed by $\mathcal{O}(q/m_{p})$, while the Dirac form factor $F_{1}$ is well described by the dipole approximation. These simplifications cease to hold at higher momentum transfers, where $F_{2}$ becomes non-negligible and the dipole form fails~\cite{Su:2024flx}. In the case of high energy DM-proton scattering with large $Q^{2}$, we therefore retain both form factors, adopting modern parameterizations that satisfy physical boundary conditions and reproduce the correct asymptotic behavior.

After applying Eq.~(\ref{eq:Lagrangian}), and Eq.~(\ref{eq:effint}), the average square amplitude of DM-proton scattering for elastic scattering case is given by
\begin{equation}
\begin{split}
    \overline{|\mathcal{M}|^{2}} = & \frac{g_{p}^{2} g_{\chi}^{2}}{\left(Q^{2}+m_{A^{\prime}}^{2}\right)^{2}} \frac{1}{4}\mathcal{X}^{\mu\nu} W_{\mu\nu} 
\end{split}
\end{equation}
with 
\begin{equation}
\begin{aligned}
    \mathcal{X}^{\mu\nu} &= 8k^{\mu } k^{\nu } +4(k^{\mu }q^{\nu }+k^{\nu }q^{\mu } )-2Q^{2} g^{\mu \nu },\\
    W_{\mu\nu} & = 4(F_{1}\left(Q^{2}\right)+F_{2}\left(Q^{2}\right)  )^{2} (2p_{\mu } p_{\nu }-p_{\mu }q_{\nu }-p_{\nu }q_{\mu } \\
    & -\frac{1}{2}Q^{2}g_{\mu \nu})+\frac{F_{2}^{2} }{m_{p}^{2} }(2m_{p}^{2}+\frac{1}{2} Q^{2}) (4p_{\mu } p_{\nu }-2p_{\mu }q_{\nu } \\
    & -2p_{\nu }q_{\mu }+q_{\mu }q_{\nu }).
\end{aligned}
\end{equation}

Then we can also obtain the differential cross-section of DM-proton elastic scattering 
\begin{equation}
\begin{aligned}
     \frac{\mathrm{d}\sigma_{\chi p}}{\mathrm{d} T_{\chi}}&= \frac{\overline{\left | \mathcal{M}  \right |^{2}}  }{32\pi m_{\chi}T\left ( T+2m_{p} \right )} \\
     &= \frac{ g_{p}^{2} g_{\chi}^{2}}{4\pi T\left ( T+2m_{p} \right )\left(Q^{2}+m_{A^{\prime}}^{2}\right)^{2}}\cdot
      [(F_{1}\left(Q^{2}\right) \\
      & +F_{2}\left(Q^{2}\right)  )^{2}\mathcal{K}_{1}+(1+\tau) F_{2}\left(Q^{2}\right)^{2}\mathcal{K}_{2}],
\end{aligned}
\end{equation}
where $\mathcal{K}_{1}=2m_{\chi }(T+m_{p} )^{2}-2 m_{\chi }T_{\chi}(T+m_{p}) -m_{p}^2T_{\chi}-m_{\chi }^2T_{\chi}+m_{\chi }T_{\chi}^2$ and $\mathcal{K}_{2}=2m_{\chi }(T+m_{p} )^{2}-2 m_{\chi }T_{\chi}(T+m_{p}) -m_{p}^2T_{\chi}$. This expression enables computation of the energy-loss rate through Eq.~(\ref{eq:energy loss}), ultimately yielding the CR cooling timescale. The complete framework now establishes quantitative predictions for CR energy loss via elastic scattering. We next extend this formalism to DIS, where the partonic degrees of freedom dominate.

Although the two form factors $F_1$, $F_2$ are independent, both contain contributions from the magnetic moment and charge distribution. Therefore, it is common to parameterize the nucleon structure $F_{1}$ and $F_{2}$ by recombining the Sachs form factors $G_{E}^{p}$ and $G_{M}^{p} $, which separately describe the electric and magnetic contributions~\cite{Sachs:1962zzc}, 
\begin{equation}
\begin{aligned}
F_{1}\left(Q^{2}\right)&=\frac{G_{E}^{p}\left(Q^{2}\right)+\tau G_{M}^{p}\left(Q^{2}\right)}{1+\tau}, \\
F_{2}\left(Q^{2}\right)&=\frac{G_{M}^{p}\left(Q^{2}\right)-G_{E}^{p}\left(Q^{2}\right)}{1+\tau},
\end{aligned}
\end{equation}
where $\tau \equiv Q^{2}/4m_{p}^{2}$. In general, the Sachs form factors are parameterized using electron--proton scattering data. In this work, we adopt the parameterization of Ref.~\cite{Kelly:2004hm} as our benchmark model. 
 
For high momentum transfers, i.e., $Q^2 \gtrsim 1~\mathrm{GeV}^2$, DM can probe the internal structure of nucleons, and DM-proton scattering is dominated by DIS. Following the collinear factorization approach detailed in Ref.~\cite{Su:2023zgr}, the DIS cross section can be expressed as an incoherent sum of DM-quark elastic scatterings, $\chi(k)+q\left(x  p\right) \rightarrow \chi\left(k^{\prime}\right)+q\left(p^{\prime}\right)$, where $x$ denotes the quark momentum fraction.

According to the Fermi's Golden Rule, the differential cross-section of DM-proton DIS can be expressed as~\cite{Su:2022wpj, Su:2023zgr}
\begin{equation}
    \begin{aligned}
\mathrm{d} \sigma_{\chi p} & =\sum_{q =u,d} \int_{0}^{1} f_q\left(x, Q^{2}\right) \mathrm{d} x \frac{(2 \pi)^{4} \delta^{4}\left(x p-q-p^{\prime}\right)}{4 \sqrt{(k \cdot x p)^{2}-\left(x m_{p} m_{\chi}\right)^{2}}} \\
&\qquad \cdot \overline{|\mathcal{M}(x)|^2} \frac{\mathrm{d}^{3} \vec{k}^{\prime}}{(2 \pi)^{3} 2 E_{k^{\prime}}} \frac{\mathrm{d}^{3} \vec{p}^{\prime}}{(2 \pi)^{3} 2 E_{p^{\prime}}} \\
& =\frac{1}{16 \pi^{2} m_{\chi} \sqrt{E_{p}^{2}-m_{p}^{2}}} \sum_{q =u,d} \int_{0}^{1} \frac{f_q\left(x, Q^{2}\right)}{x} \mathrm{d} x \\
&\qquad \cdot \overline{|\mathcal{M}(x)|^2} \delta^{4}\left(x p-q-p^{\prime}\right) \frac{\mathrm{d}^{3} \vec{k}^{\prime} \mathrm{d}^{3} \vec{p}^{\prime}}{4 E_{k^{\prime}} E_{p^{\prime}}}.
\end{aligned}
\end{equation}
Then we can rewrite the $\delta$ function and obtain
\begin{equation}
\begin{aligned}
    \frac{\mathrm{d} \sigma_{\chi p}}{\mathrm{d} T_{\chi }} &= \frac{\left|\vec{k}^{\prime}\right|}{16 \pi m_{\chi} Q^{2} \sqrt{E_{p}^{2}-m_{p}^{2}}} \\
    &\quad \times \sum_{q =u,d}\int_{\cos \theta_{\min }}^{1} f_q\left(x, Q^{2}\right) \overline{|\mathcal{M}(x)|^{2}} \mathrm{~d} \cos \theta, 
\end{aligned}
\end{equation}
where the momentum fraction $x = Q^2 /(-2 p\cdot q)$ is the function of $Q^2 = 2 m_{\chi} T_{\chi}$ and $-2 p\cdot q = 2 |\vec{p}| |\vec{k}^\prime| \cos \theta -2 E_p T_{\chi}$. The lower limit $\cos\theta_{\min}$ is determined by the kinematic constraint $0 < x \le 1$. The invariant mass $W$ of the final-state hadronic system satisfies $W^2 = m_p^2 - 2p\cdot q - Q^2$. The intermediate nucleon-resonance region corresponds approximately to $m_p+m_\pi\leq W\lesssim2\,\mathrm{GeV}$, where $m_\pi$ is the pion mass~\cite{Christy:2007ve}. Due to the energetic protons in the cosmic rays, the contribution of the DIS region dominates over that of the intermediate resonance region in  our study. The DIS contribution vanishes when no angular interval satisfies the kinematic constraints or when the adopted $Q^2$ cut is not satisfied. In this work, we focus on the partonic contributions from up and down quarks. Their parton distribution functions (PDFs), $f_q(x,Q^2)$, are taken from the results of CT10nlo as implemented in the LHAPDF package~\cite{Lai:2010vv, Buckley:2014ana}. In this frame, the spin-averaged squared amplitude for DM-quark scattering is
\begin{equation}
\begin{split}
    \overline{|\mathcal{M}(x)|^{2}} = & \frac{g_{\chi}^{2} g_{q}^{2}}{\left(Q^{2}+m_{A^{\prime}}^{2}\right)^{2}} \frac{1}{4}\mathcal{L}^{\mu\nu} H_{\mu\nu}. 
\end{split}
\end{equation}
where dark matter current tensor $\mathcal{L}^{\mu\nu}$ and hadronic current tensor $\mathcal{H}^{\mu\nu}$ are
\begin{equation}
\begin{aligned}
    \mathcal{L}^{\mu\nu} &= 8k^{\mu } k^{\nu } +4(k^{\mu }q^{\nu }+k^{\nu }q^{\mu } )-2Q^{2} g^{\mu \nu },\\
    H_{\mu\nu} &= 8 x p_{\mu} \cdot x p_{\nu}-4\left(x p_{\mu} \cdot q_{\nu}+q_{\mu} \cdot x p_{\nu}\right)+2 q^{2} g_{\mu \nu}.
\end{aligned}
\end{equation}
Therefore,

\begin{equation}
\begin{aligned}
\overline{|\mathcal{M}(x)|^{2}} 
&=  \frac{g_{\chi}^{2} g_{q}^{2}}{\left(Q^{2}+m_{A^{\prime}}^{2}\right)^{2}} 
\left[ 16\left(x m_{\chi} E_{p}\right)^{2} - 8 x m_{\chi} E_{p} Q^{2} \right. \\
&\quad \left. -4 Q^{2} m_{\chi}^{2} -4 Q^{2} x^{2} m_{p}^{2} +2\left(Q^{2}\right)^{2} \right].
\end{aligned}
\end{equation}

\begin{figure*}[t] 
    \centering
    \begin{minipage}[t]{0.48\linewidth}
        \centering
        \includegraphics[width=\textwidth]{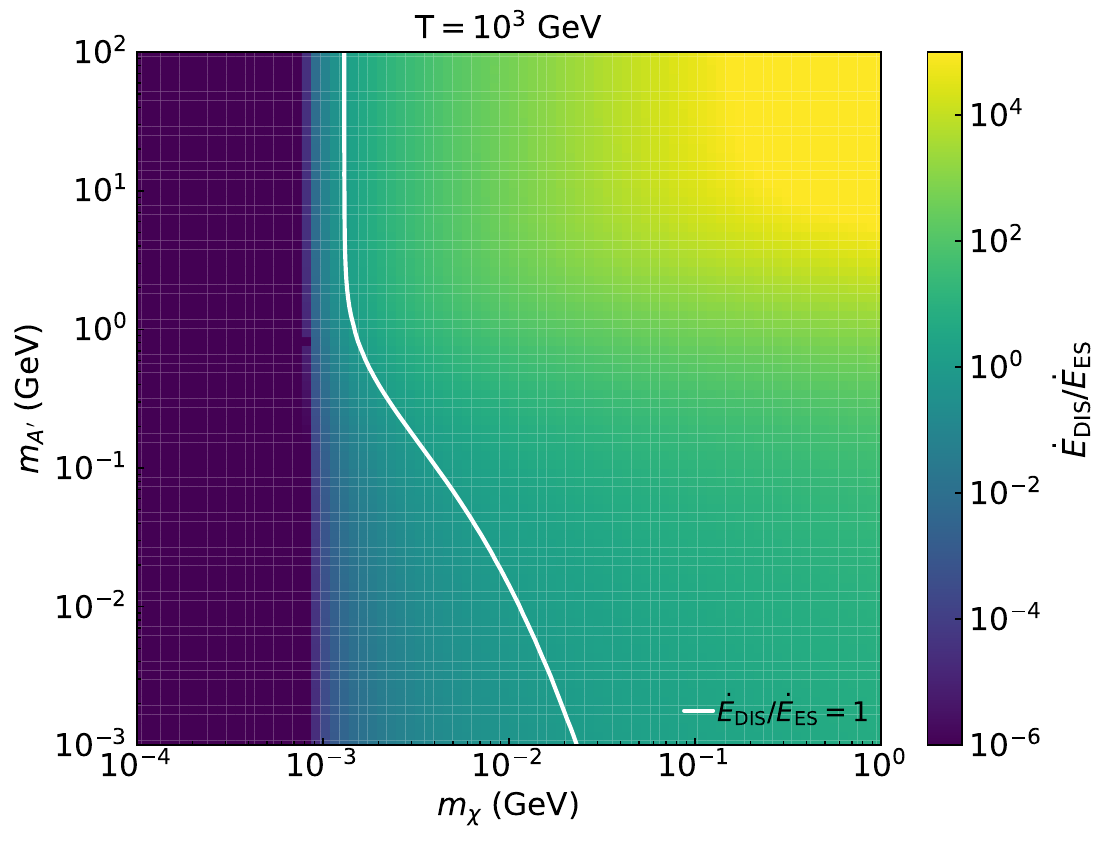}
    \end{minipage}
    \hfill
    \begin{minipage}[t]{0.48\linewidth}
        \centering
        \includegraphics[width=\textwidth]{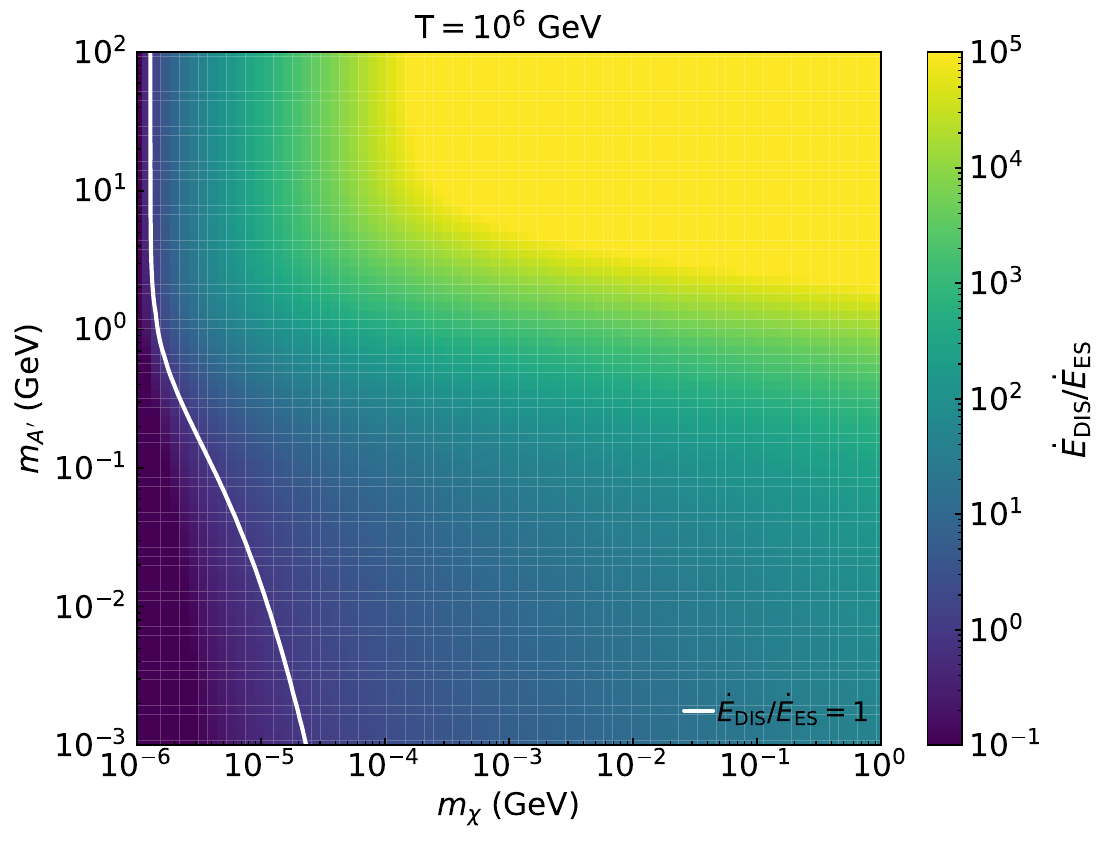}
    \end{minipage}
    \caption{Ratio $\dot{E}_{DIS}/\dot{E}_{ES}$ as a function of DM mass $m_{\chi}$ and mediator mass $m_{A'}$ for cosmic ray energies, $10^{3} ~\mathrm{GeV}$ (\textit{Left panel}) and $10^{6}~\mathrm{GeV}$ (\textit{Right panel}). The white contour denotes $\dot{E}_{DIS}/\dot{E}_{ES}=1$, where DIS and elastic scattering contribute equally to energy loss rate. Both panels show maximal ratio in the upper-right quadrant where $m_{A'}$ and $m_{\chi}$ are simultaneously large.}
    \label{colormap}
\end{figure*}
 
\textbf{\textit{Result and Discussion:}} 
To compare the contributions of elastic and inelastic scattering processes to the CR cooling rate, we estimate the energy loss for both process. Fig.~\ref{colormap} shows the ratio of inelastic to elastic scattering energy loss rate,  $\dot{E}_{DIS}/\dot{E}_{ES}$, as a function of the mediator mass $m_{A'}$ and DM mass $m_\chi$ for CR energies of $T = 10^3~\mathrm{GeV}$ (\textit{Left panel}) and $T = 10^6~\mathrm{GeV}$ (\textit{Right panel}). The white contour line corresponds to the case where the elastic and inelastic scattering energy loss rates  contribute equally, i.e., $\dot{E}_{DIS}/\dot{E}_{ES}=1$. It is evident that both $m_{A'}$ and $m_\chi$ critically affect the scattering cross-section and thus the CR energy loss rates. The $m_{A'}$ modulates the cross-section through its propagator $(Q^2 + m_{A'}^2)^{-2}$, while $m_\chi$ determines the momentum transfer threshold $Q^{2}=2m_{\chi}T_{\chi}$.

Our two specific CR kinetic energies, \(T = 10^{3}\)~GeV and \(T = 10^{6}\)~GeV, are selected to probe two typical regimes of CR-DM scattering: the lower energy corresponds to the regime where elastic scattering makes a non-negligible contribution to CR-DM energy transfer, while the higher energy corresponds to the regime where DIS makes a much larger contribution. These two benchmarks allow us to systematically compare the impact of elastic and DIS channels on the total CR cooling rate, with results for additional CR energies presented in Appendix A.

We observe that for low DM masses, the momentum transfer is low. This results in a small energy loss rate for the DIS process, allowing elastic scattering to dominate the total energy loss. A larger $m_{\chi}$ increases the momentum transfer, thereby enhancing the relative contribution of DIS to the energy loss rate. Consequently, the ratio $\dot{E}_{\mathrm{DIS}}/\dot{E}_{\mathrm{ES}}$ increases systematically. An abrupt kinematic cutoff occurs near $m_{\chi} \approx 10^{-3}$ GeV for $10^{3}$ GeV CRs and near $m_{\chi} \approx 10^{-6}$ GeV for $10^{6}$ GeV CRs. This cutoff corresponds to the point where the DIS contribution becomes equal to that of elastic scattering. This transition arises when the typical momentum transfer approaches $Q^2 \sim 1$ GeV$^2$, the threshold above which DIS becomes kinematically favorable. Furthermore, for higher CR energies (e.g., $T=10^6$ GeV), the energy loss rate via DIS is significant even at lower DM masses, in contrast to the low-energy CR case (e.g., $T=10^3$ GeV). For sufficiently large DM masses (specifically, $m_{\chi}>0.1$ GeV for $T=10^3$ GeV and $m_{\chi}>10^{-4}$ GeV for $T=10^6$ GeV), the energy loss from DIS exceeds that from elastic scattering by many orders of magnitude. Our findings are complementary to and consistent with those in Ref.~\cite{Su:2023zgr}, a point further verified in Appendix B through analysis of the energy-dependent inelastic scattering cross-section.

The DIS channel provides a substantial contribution to the energy-transfer rate over its kinematically allowed phase space. While Fig.~\ref{colormap} confirms DIS dominance over elastic scattering persists across mediator masses, heavier mediators ($m_{A'}>1$ GeV) markedly amplify this prevalence by suppressing the $1/Q^4$ enhancement characteristic of elastic scattering at low momentum transfers. Consequently, in AGN environments like NGC 1068 with high CR kinetic energies, DIS contributes the majority of the total DM-induced energy loss for massive dark sector carriers, creating distinctive cooling signatures inaccessible to standard astrophysical processes.

\begin{figure}[]
\centering
\includegraphics[width=8cm]{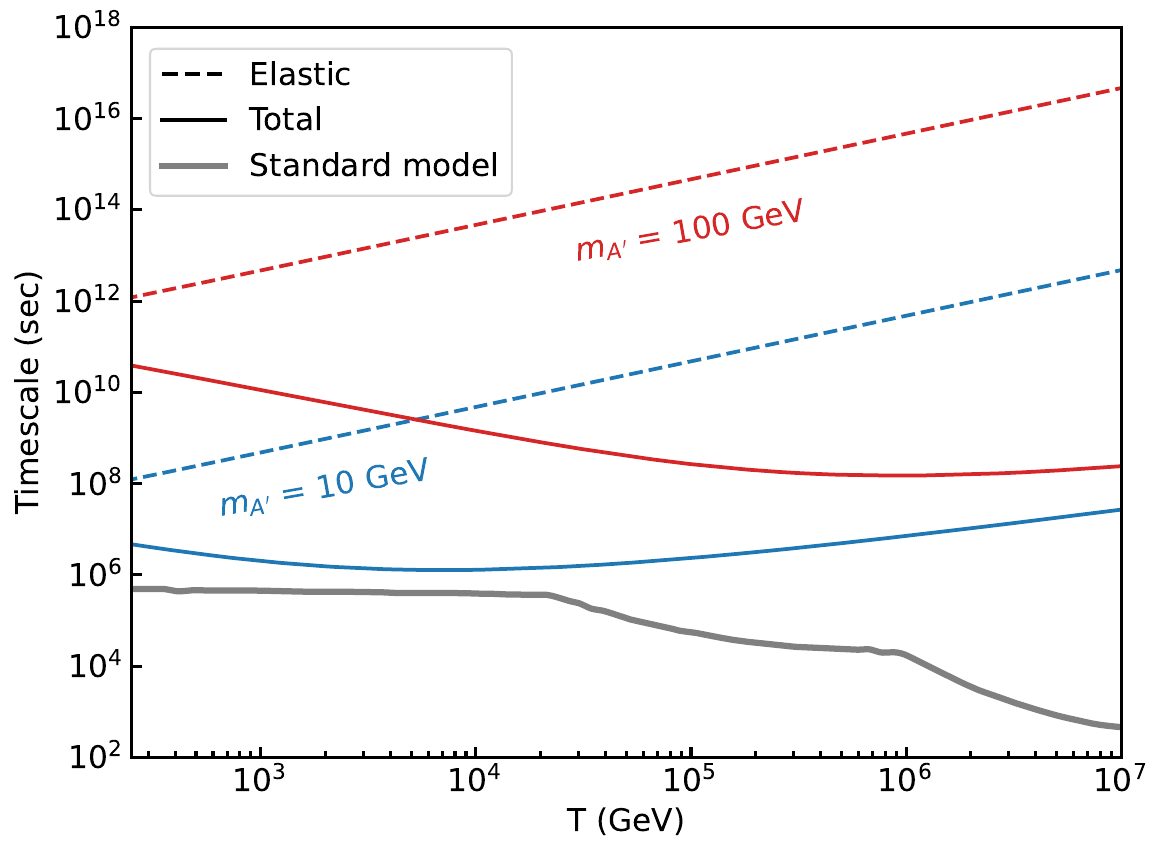}
\caption{Cooling timescale for CRs interacting with DM as a function of CR kinetic energy. The dashed lines correspond to DM-proton elastic scattering processes, and the solid lines correspond to the combination of both elastic and deep inelastic scattering processes. Additionally, the gray line indicates CR cooling via standard model processes. Here, we fix the DM mass $m_{\chi}=0.1~\mathrm{GeV}$.}
\label{fig:timescale}
\end{figure}

To quantify these potential DM-induced energy loss and derive constraints, we compare the DM cooling timescale $t_{\chi p}$ with the cooling timescale from SM processes in NGC 1068, denoted as $t_{\rm SM}$. The gray line in Fig.~\ref{fig:timescale}  represents this total SM cooling timescale, derived from multi-messenger modeling that reproduces the observed neutrino and gamma-ray spectra. It includes the dominant SM energy-loss processes: inelastic \(pp\) collisions (dominant at \(T \lesssim 10^{4}\)~GeV), Bethe-Heitler pair production (\(10^{4} \lesssim T \lesssim 10^{6}\)~GeV), and \(p\gamma\) interactions at higher energies\cite{Murase:2019vdl, Fiorillo:2024akm, Fiorillo:2025ehn}.

Fig.~\ref{fig:timescale} presents CR cooling timescales in NGC 1068 for fixed couplings $g_{\chi} = g_{p} = 0.1$, comparing elastic scattering (Elastic, dashed lines) with combined elastic and DIS contributions (Total, solid lines). Blue and red curves correspond to a mediator mass of $m_{A'}=10~\mathrm{GeV}$, and $100~\mathrm{GeV}$, both are compared with the SM cooling rate (solid gray line). As illustrated in Fig.~\ref{fig:timescale}, for light DM ($m_{\chi}=0.1~\mathrm{GeV}$) coupled to a mediator with $m_{A'} = 10~\mathrm{GeV}$, the cooling timescale exceeds that of the SM across the entire range of kinetic energy $T$. The scenario with $m_{A'} = 100~\mathrm{GeV}$ results in a further prolonged cooling timescale compared to both the SM and the $m_{A'}=10~\mathrm{GeV}$ case. Notably, for both elastic and DIS processes, the timescale universally shortens with decreasing $m_{A'}$.

Furthermore, the systematic inclusion of DIS universally reduces the cooling timescales across the full kinetic energy range. At low energies, the small momentum transfer $Q^2$ diminishes differences between the DIS and elastic scattering timescales, while the disparity increases significantly at high energies. For the case with $m_{A'} = 10~\mathrm{GeV}$, a distinctive dip emerges at low energies due to suppressed partonic structure access from both large mediator mass and limited $Q^2$. However, at high energies (e.g., $10^7~\mathrm{GeV}$), DIS exhibits a marked reduction in timescale relative to elastic scattering. Although the ratio $\dot{E}_{DIS}/\dot{E}_{ES}$ increases with $m_{A'}$, for sufficiently heavy mediators ($m_{A'}\gtrsim 10~\mathrm{GeV}$), the cooling timescale approaches that of the $m_{A'} = 10~\mathrm{GeV}$ case for high-energy CRs, as shown in Fig.~\ref{fig:timescale}. Given the enhanced cooling efficacy, characterized by a pronounced DIS contribution at high energies and shorter overall timescales, we select the $m_{A'} = 10~\mathrm{GeV}$ scenario for detailed investigation. The results for other mediator masses are discussed in Appendix C.

To derive constraints on the gauged $U(1)$ fermionic DM model, we apply the CR cooling condition $t_{\chi p} \leq C t_{\mathrm{SM}}$ to exclude parameter space where cooling is too rapid. Here, $t_{\mathrm{SM}}$ denotes the CR proton cooling timescale via SM processes. Systematic uncertainties in $t_{\rm SM}$, arising from the precise CR injection spectrum and the astrophysical conditions of the emission region, are encapsulated in the tolerance factor $C$~\cite{Herrera:2023nww}. For NGC 1068, where the proton luminosity $L_{\mathrm{p}}$ ranges from $10^{43}~erg~s^{-1}$ to $10^{44}~erg~s^{-1}$, the value of $C$ can vary between 0.1 and 1. Our conservative choice of \(C = 0.1\) implies we only exclude parameter space where the DM-induced cooling would be at least ten times more efficient than the SM baseline, ensuring robustness against potential overestimation of $t_{\rm SM}$.

\begin{figure}[]
\centering
\includegraphics[width=8cm]{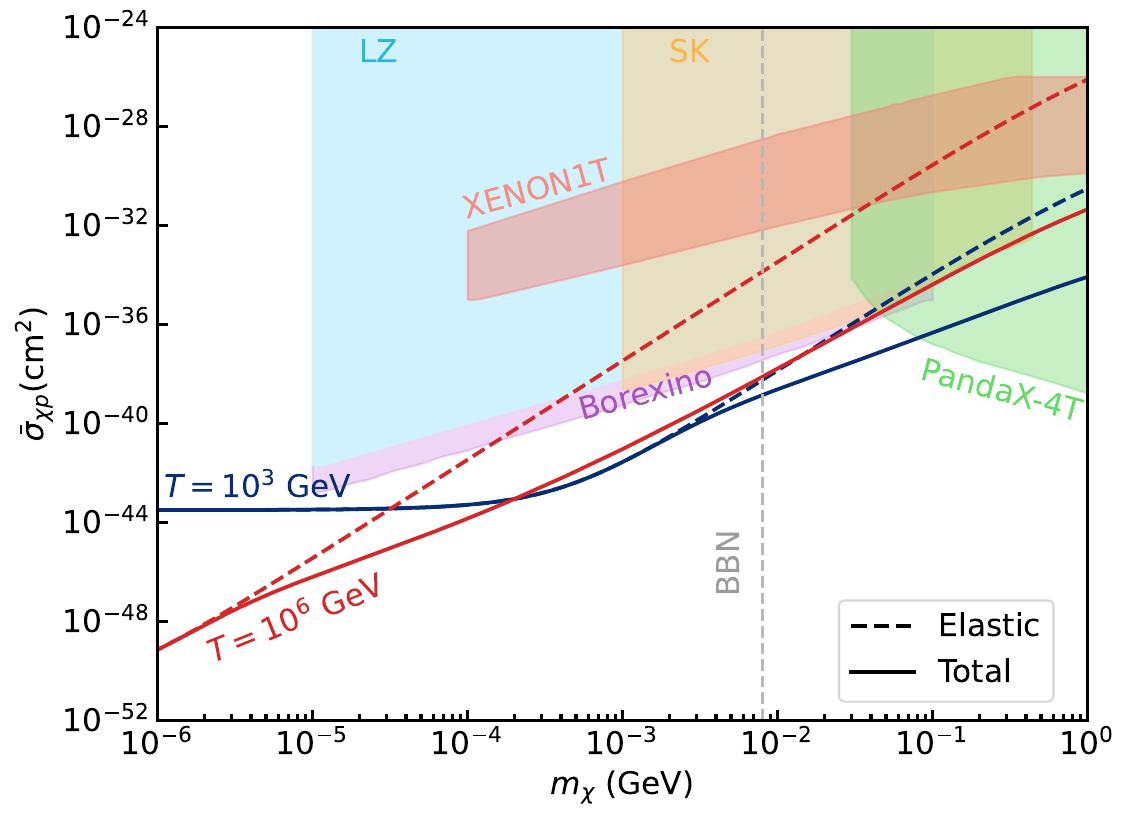}
\caption{Constraint on the DM-proton elastic and deep inelastic scattering cross-section from the CR cooling effect, with a mediator mass of $m_{A'} = 10 \text{ GeV}$. Solid curves represent the limits including both elastic and DIS, while dashed curves correspond to elastic-only limits. Red and blue colours indicate kinetic energy of $T = 10^6 \text{GeV}$ and $T = 10^3 \text{GeV}$, respectively. The bounds from XENON1T (light red)~\cite{Dent:2019krz}, LZ (sky blue)~\cite{Wang:2025ztb, LZ:2024zvo}, SK (yellow)~\cite{Bell:2023sdq}, Borexino (purple)~\cite{Wang:2025ztb, Borexino:2013bot}, PandaX-4T (halo DM, Migdal effect, green)~\cite{PandaX:2023xgl}, and BBN (gray)~\cite{Giovanetti:2021izc} are shown.}
\label{fig:limit}
\end{figure}

In order to facilitate comparison with the existing experiments, a reference cross-section for DM-proton scattering is defined in a model-independent way as $\bar{\sigma}_{\chi p}=(g_{p}^{2} g_{\chi}^{2} \mu_{\chi p}^{2})/(\pi m_{A^{\prime}}^{4})$ where $\mu_{\chi p}$ is the reduced DM-proton mass. Fig.~\ref{fig:limit} shows the constraints for the DM-proton cross-section as a function of DM mass. It is evident that for low CR energies ($T = 10^3$ GeV) and low DM masses ($m_\chi < 10^{-3}$ GeV), the momentum transfer rate is low. Consequently, the energy loss rate and cooling timescale for elastic and DIS are approximately equal. However, for comparatively higher DM masses ($m_\chi > 10^{-3}$ GeV), the enhanced momentum transfer causes DIS to dominate over elastic scattering. As a result, constraint on the DIS cross-section is approximately three orders of magnitude stronger than that for the elastic scattering case.

Furthermore, at higher CR energies ($T = 10^6$ GeV), the energy loss rate from DIS is consistently greater than that from elastic scattering for DM masses ranging from $10^{-6}$ GeV to $1$ GeV. Consequently, the resulting DIS cross-section limits exceed those from elastic scattering by an average of at least three orders of magnitude throughout this mass range. Fig.~\ref{fig:limit} compares our CR-cooling constraints with the CRDM bounds obtained by reinterpreting XENON1T data~\cite{XENON:2018voc,Dent:2019krz}. The upper boundary of the XENON1T exclusion region reflects the approximate Earth-attenuation treatment in Ref.~\cite{Dent:2019krz}: at sufficiently large cross sections, DM loses too much energy before reaching the underground detector to produce detectable recoils. We also show the LUX-ZEPLIN (LZ) bounds on blazar-boosted dark matter obtained in Ref.~\cite{Wang:2025ztb} recasting the LZ data reported in Ref.~\cite{LZ:2024zvo}. These constraints also surpass bounds from direct detection experiments such as PandaX-4T (via Migdal effect)~\cite{PandaX:2023xgl}, as well as constraints from neutrino experiments including Super-Kamiokande (SK)~\cite{Super-Kamiokande:2022ncz} and Borexino~\cite{Wang:2025ztb}. Our results further encompass the vertical bounds set by the  BBN~\cite{Giovanetti:2021izc}. This suggests that for a heavy vector mediator, CR cooling via DM-proton DIS provides constraints that are competitive with existing direct detection bounds, particularly in the sub-GeV DM mass region. Therefore, we find that the cooling of CR protons via DM interactions provides a powerful and novel probe of DM properties. 

We note that the DM spike profile and the constraints derived in this work are affected by several key sources of systematic uncertainty. The spike’s survival critically depends on a quiescent galactic merger history, as major mergers can efficiently erase it~\cite{Meighen-Berger:2025hrq}. While NGC 1068 is suggested to have a non-merger history, supporting the possibility of an intact spike, other mechanisms could further modify the profile. DM annihilation or self-interactions would reduce the spike's density and weaken our limits~\cite{Chiang:2019zjj,Alvarez:2020fyo}, and baryonic feedback from stars and other matter may significantly alter the spike profile~\cite{Papastergis_2016}.

In addition, the DM spike profile itself is subject to substantial theoretical uncertainty. When the inner slope $\gamma$ of the initial NFW profile varies between $0.6$ and $1.4$, the resulting spike slope $\gamma_{\mathrm{sp}}$ changes correspondingly. This is obtained via adiabatic compression, leading to variations in $\rho_{\chi}$ by up to two to three orders of magnitude as discussed in Ref.~\cite{Herrera:2023nww}. Since the energy loss rate $\dot{E}_{\chi p}$ is directly proportional to $\rho_\chi$ (Eq.~\ref{eq:energy loss}), the cooling time $t_{\chi p}$ scales inversely with $\rho_\chi$ and therefore varies by a comparable factor. Accounting for this uncertainty is essential when interpreting the timescale constraints, and incorporating such astrophysical effects will be critical for future refinements. Looking forward, multi-messenger observations will be required to solidify these constraints and provide profound insights into the microphysical nature of DM and its dynamics around SMBHs.

Beyond the above uncertainties in the DM spike profile, we also note that a more sophisticated derivation has recently been proposed in Ref.~\cite{Kantzas:2025huu}. In their paper, the authors consider cosmic-ray survival during acceleration and escape from the host galaxy separately, thereby providing a more detailed treatment. The approach further systematically incorporates uncertainties in key astrophysical parameters, including the magnetic field strength, acceleration efficiency, and the location of the emission region. While this method is developed for blazar environments dominated by leptonic processes, it can in principle be adapted to AGN-like environments such as NGC~1068. A more comprehensive treatment that incorporates these uncertainties would further refine our constraints on $\sigma_{\chi p}$. Nevertheless, while the exclusion limits may be subject to such improvements, our main conclusion regarding the importance of the DIS processes remains robust.

\textbf{\textit{Conclusion:}}
Cosmic-ray cooling in active galactic nuclei has emerged as a powerful probe of interactions between dark matter and standard model particles. In this work, we have demonstrated that deep inelastic scattering critically shapes cosmic-ray energy loss within active galactic nuclei dark matter spikes. Our analysis of NGC 1068 shows that for high-energy (TeV-PeV) protons, deep inelastic scattering dominates energy loss, facilitating efficient energy transfer to dark matter via hadronic fragmentation. For heavier mediators ($m_{A'} \gtrsim 1~\mathrm{GeV}$) at these energies, the deep inelastic scattering process enhances the cross-section, shortening the cosmic-ray cooling timescale by orders of magnitude compared to elastic scattering. Consequently, we derive dramatically strengthened exclusion limits on scattering cross-section of sub-GeV dark matter. These results establish novel constraints that surpass current limits from direct detection and boosted dark matter searches, opening a new avenue for probing dark matter interactions in extreme astrophysical environments.

\section*{Acknowledgments}

This work is supported by the National Natural Science Foundation of China (NNSFC) under grants No. 12335005, No. 12275134, No. 12575118 and the Special funds for postdoctoral overseas recruitment, Ministry of Education of China. The authors gratefully acknowledge the valuable discussions and insights provided by the members of the China Collaboration of Precision Testing and New Physics.

\textit{Note added --} During the completion of this work, an independent analysis of DM-proton DIS effects at AGN sources and detectors appeared as a preprint~\cite{Gustafson:2025dff}. Their study focuses on CR-boosted inelastic DM scenarios in direct detection and neutrino experiments. In contrast, our work provides a more systematic treatment of DM-proton DIS for CR cooling in AGN, extending to both heavy and light mediators, DM annihilation, and a broad range of CR energies. Furthermore, we incorporate more sophisticated form factors and address the transition between elastic scattering and DIS across momentum transfers, key aspects for accurately modeling DIS effects.

\bibliography{refs}

\clearpage
\appendix


\section{Appendix A: Variation of exclusion limits with cosmic-ray kinetic energy}

\begin{figure*}[h!]
    \centering
    \begin{tabular}{cc}
        \includegraphics[width=0.48\textwidth]{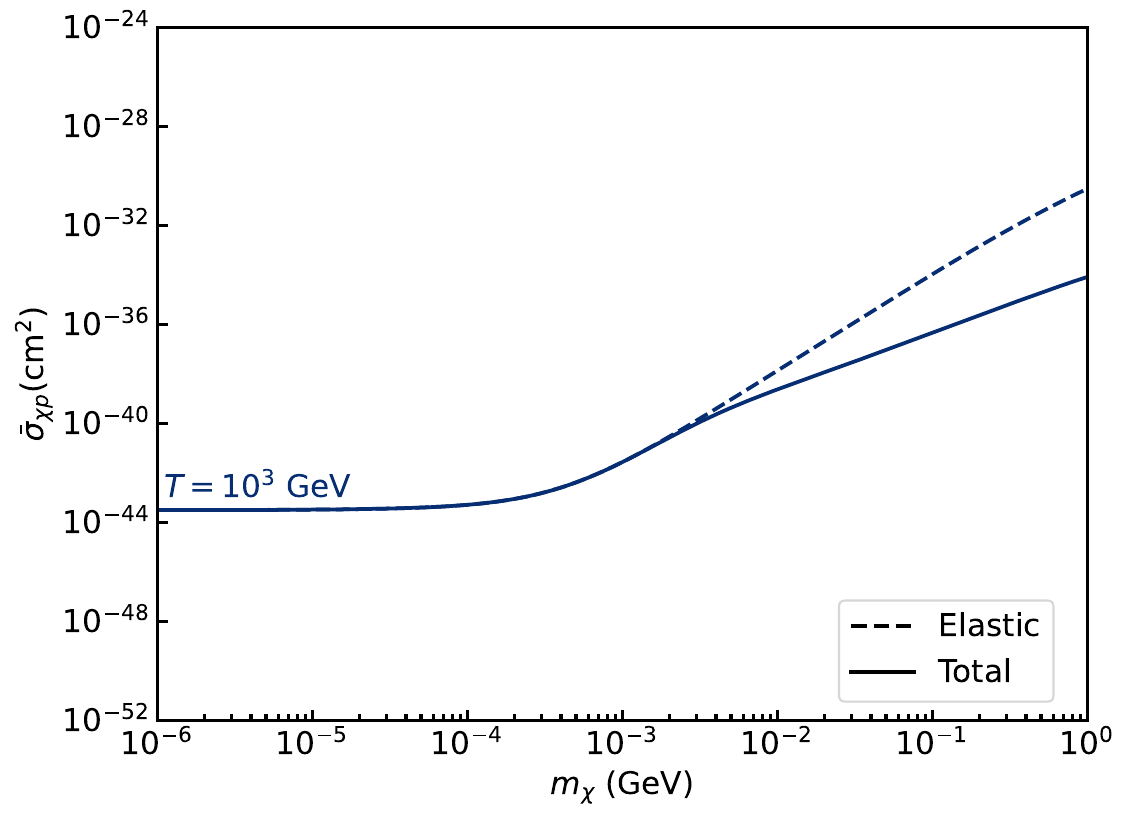} &
        \includegraphics[width=0.48\textwidth]{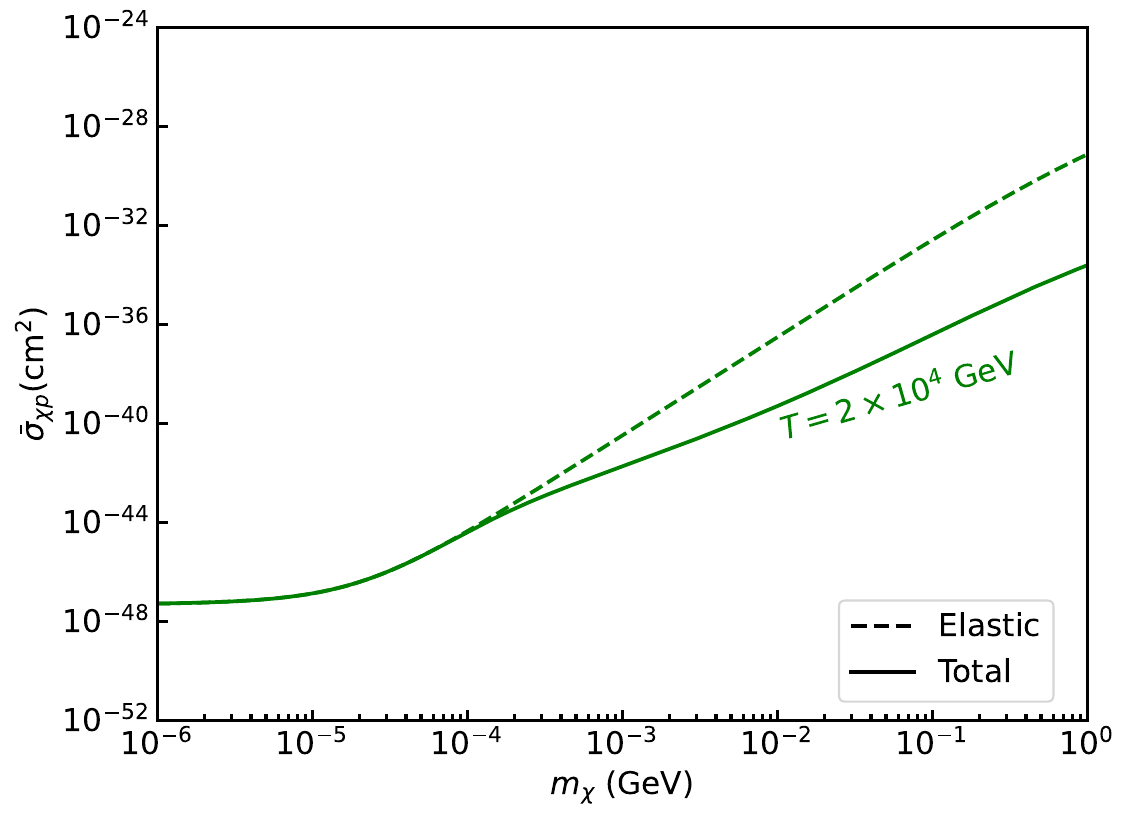} \\
        
        \vspace{0.2cm} & \vspace{0.2cm} \\
        
        \includegraphics[width=0.48\textwidth]{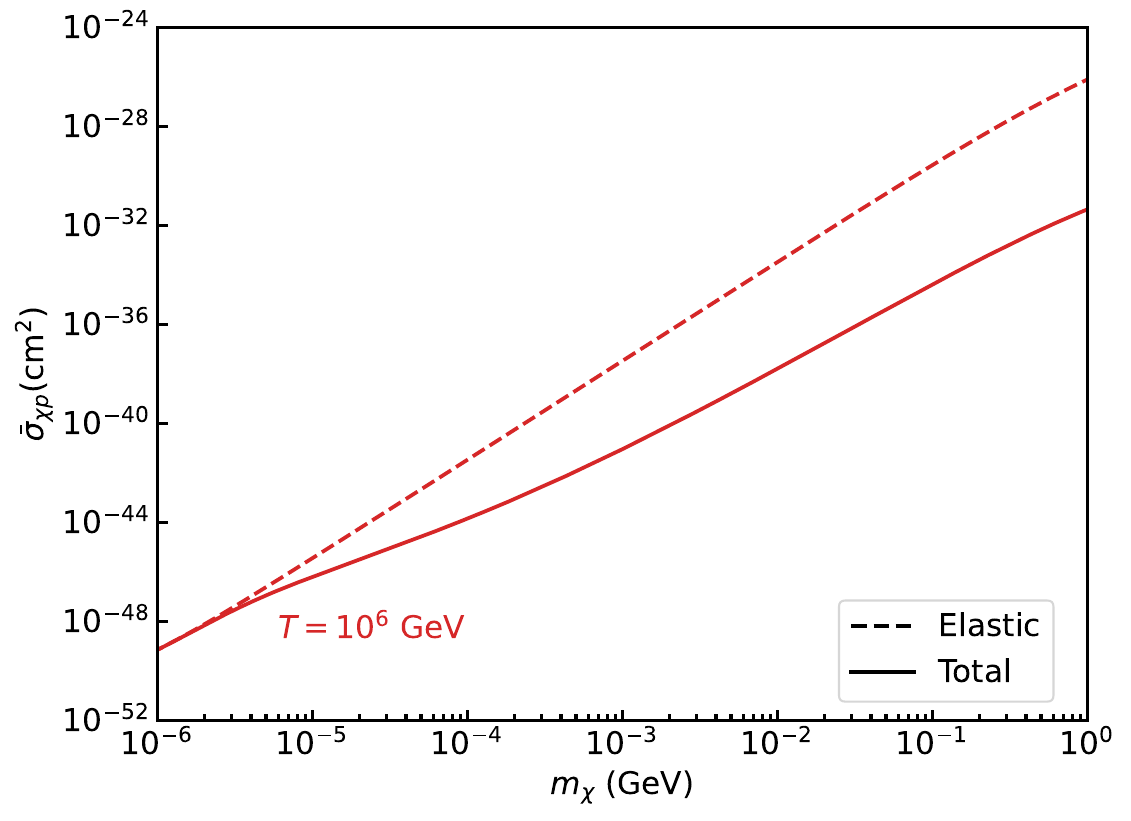} &
        \includegraphics[width=0.48\textwidth]{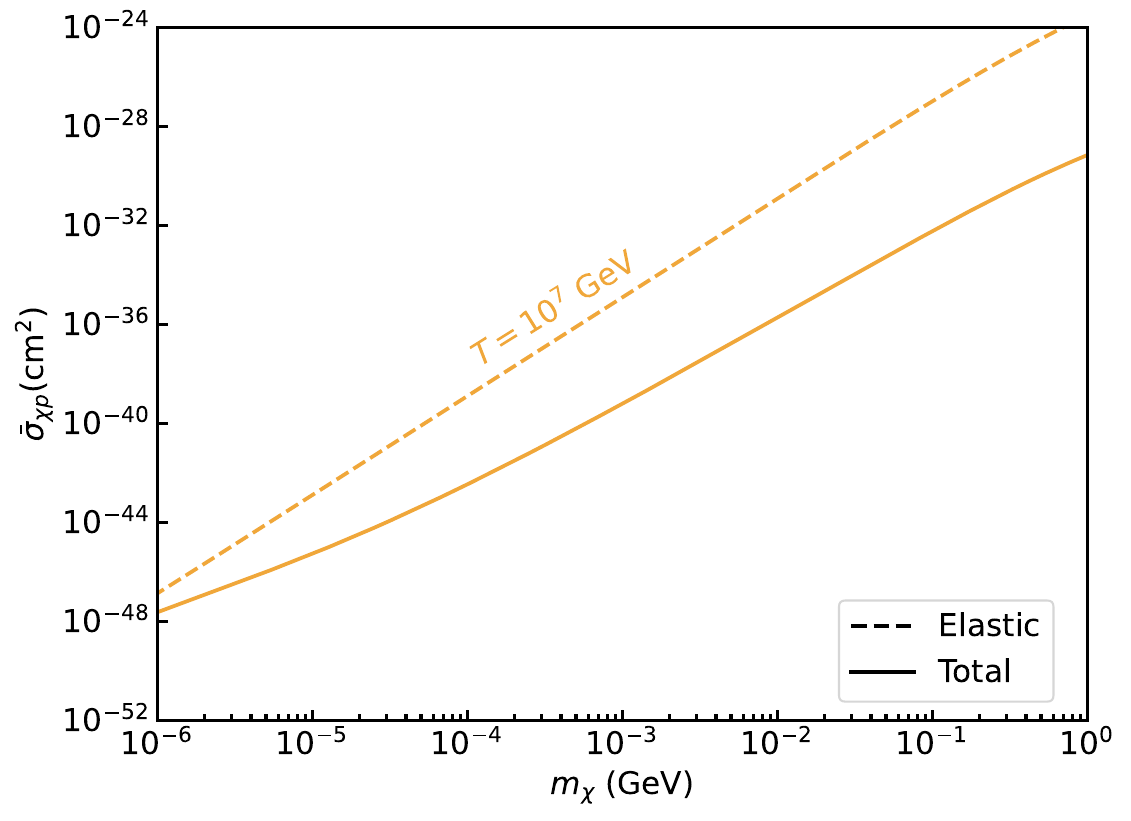}
    \end{tabular}
    \caption{Exclusion limits on the DM-proton cross-section as a function of DM mass for four CR proton kinetic energies: $T=10^{3}$ GeV (blue), $2 \times 10^{4}$ GeV (green), $10^{6}$ GeV (red), and $10^{7}$ GeV (yellow), assuming a mediator mass $m_{A'}=10$ GeV. Dashed lines correspond to limits from elastic scattering only, while solid lines represent the combined constraints from both elastic and deep-inelastic scattering (Total).}
    \label{fig:energyrange}
\end{figure*}

In the main text, we present and compare exclusion limits for two CR proton kinetic energies: $T=10^3~\mathrm{GeV}$ and $T=10^6~\mathrm{GeV}$. To verify the robustness of our constraints, we have computed the exclusion limits across a broader range of CR energies. The full set of energies ($T=10^3, 2\times 10^4, 10^6, 10^7~\mathrm{GeV}$) are selected to span the transitions between the dominant SM energy-loss processes for CR protons in NGC 1068: inelastic $pp$ collisions dominate below $\sim 10^4$~GeV, Bethe-Heitler pair production in the range $10^4 \lesssim T \lesssim 10^6$~GeV, and $p\gamma$ interactions at higher energies. Fig.~\ref{fig:energyrange} shows the limits on the DM-proton cross-section assuming our mediator mass $m_{A'}=10 \mathrm{GeV}$.

The results demonstrate a systematic evolution of the constraints with CR energy. In the very light DM mass region ($m_{\chi} \sim 10^{-6}$ GeV), the limits become progressively stronger as $T$ increases from $10^3$ to $10^6\mathrm{GeV}$. This strengthening is an outcome of the growing dominance of the DIS process at higher energies, where the larger accessible momentum transfer $Q^2$ enables more efficient energy transfer to the DM via hadronic fragmentation. Furthermore, the separation between the limits derived from elastic scattering only and those from total contributions widens with increasing $T$, reflecting the increasing importance of DIS. This systematic analysis confirms that the transition in constraint strength shown in the main text (e.g., in Fig.~\ref{fig:limit}) is a robust feature reflecting the underlying scattering dynamics across the relevant energy range.

\section{Appendix B: Variation of deep inelastic scattering cross section and energy loss rate with CR energy}

This section details how the deep inelastic scattering (DIS) cross section and energy loss rate for dark matter-cosmic ray (CR) proton collisions vary with CR energy. We discuss the enhancement of these quantities with both mediator mass and cosmic ray energy.

\begin{figure*}[h!] 
    \centering
    \begin{minipage}[t]{0.48\linewidth}
        \centering
        \includegraphics[width=\textwidth]{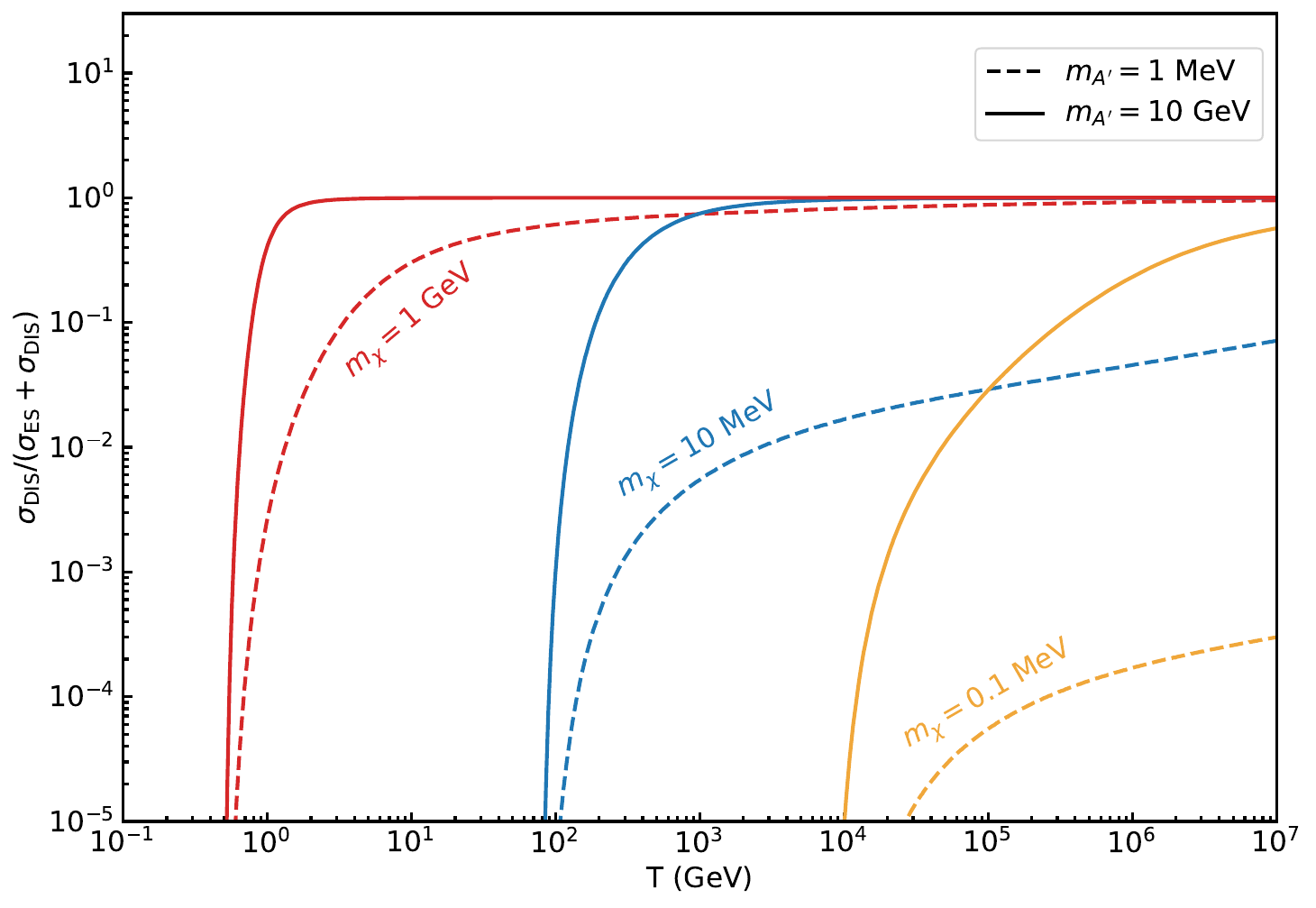}
    \end{minipage}
    \hfill
    \begin{minipage}[t]{0.48\linewidth}
        \centering
        \includegraphics[width=\textwidth]{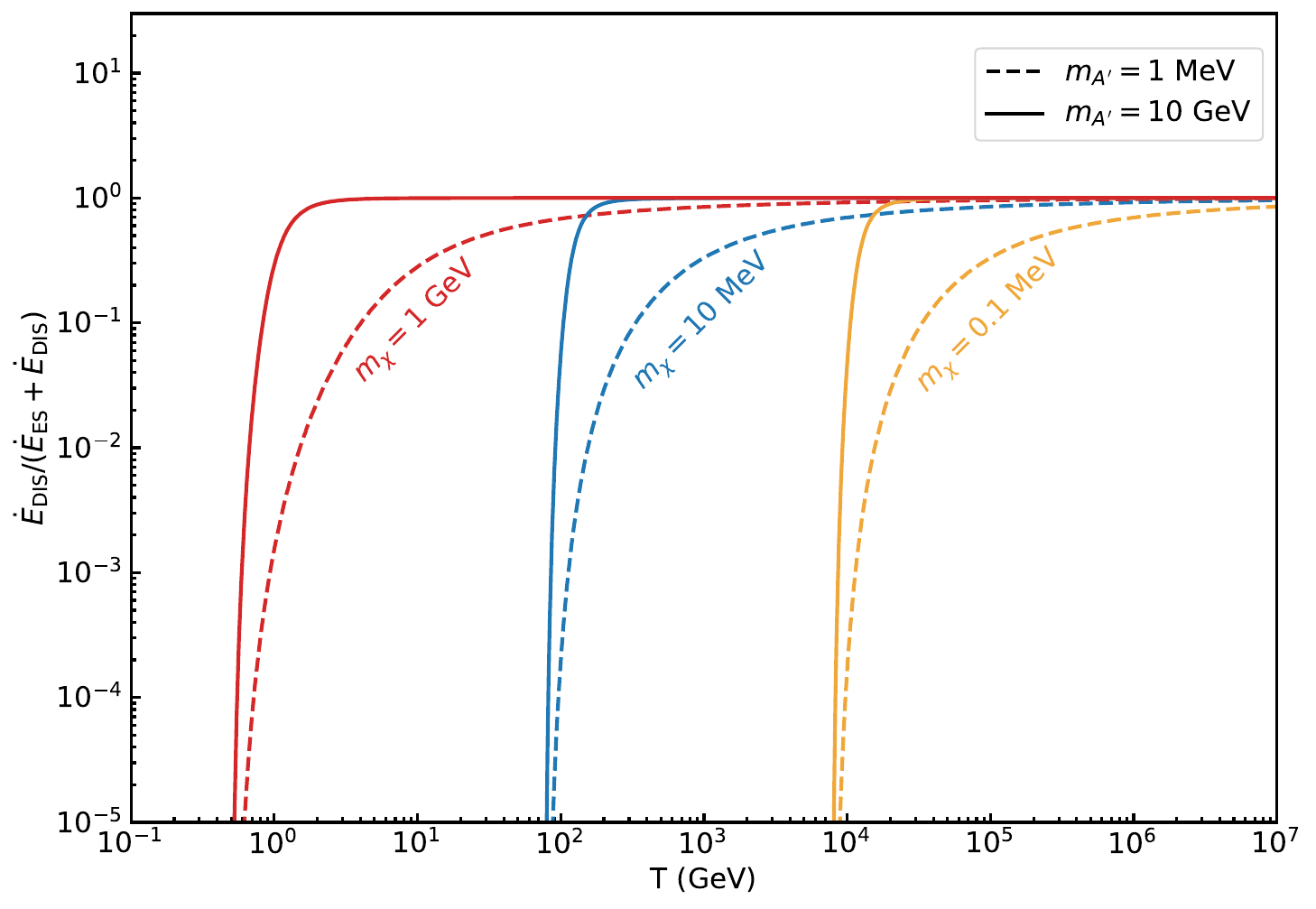}
    \end{minipage}
    \caption{(Left) Cross-section ratio, $\sigma_{\rm DIS}/(\sigma_{\rm ES}+\sigma_{\rm DIS})$, versus cosmic ray kinetic energy $T$ for DM masses: 0.1 MeV (yellow), 10 MeV (blue), 1 GeV (red). Solid curves represent the heavy mediator case ($m_{A'}=10~\mathrm{GeV}$), while dashed curves show the light mediator scenario ($m_{A'}=1~\mathrm{MeV}$). (Right) Corresponding energy loss rate ratio versus CR kinetic energy, using identical color and line-style coding.}
    \label{fig:6}
\end{figure*}

Our analysis of the DIS cross-section ratio, $\sigma_{DIS}/(\sigma_{ES}+\sigma_{DIS})$, reveals critical mediator mass dependencies across CR kinetic energies $T$ and DM masses ($m_{\chi}=0.1~\mathrm{MeV}, 10~\mathrm{MeV}, 1~\mathrm{GeV}$). For the heavy mediator ($m_{A'}=10 ~\mathrm{GeV}$, solid lines in the left panel of Fig.~\ref{fig:6}), the ratio approaches unity at high energies, confirming DIS dominance when $Q^{2}>1 \mathrm{GeV} ^{2}$. This predominance abruptly collapses at low energies where $Q^{2}<1 \mathrm{GeV} ^{2}$, the kinematic threshold for DIS suppression. Conversely, for light DM masses (e.g., $0.1~\mathrm{MeV}$ and $10~\mathrm{MeV}$), the light mediator case ($m_{A'}=1~\mathrm{MeV}$, dashed lines) exhibits persistently low ratios across all energies, indicating that elastic scattering remains the dominant process. This result is consistent with the findings in Ref.~\cite{Su:2023zgr} (see Fig.~4), which reports subdominant DIS contributions relative to elastic scattering ones for light DM and a mediator mass of $m_{A'}=0.3~\mathrm{GeV}$ in the DM kinetic energy range of $0.1$-$10~\mathrm{GeV}$.

Crucially, the CR energy-loss rate in Eq.~\ref{eq:energy loss} weights the differential cross section by the transferred energy \(T_\chi\). Consequently, the relative contributions to the energy-loss rate need not follow the corresponding total cross-section ratios. As shown in the right panel of Fig.~\ref{fig:6}, the DIS contribution becomes dominant in the parameter regions where the energy-loss ratio approaches unity, including the light-mediator benchmarks at sufficiently high CR energies.

\section{Appendix C: Dependence of exclusion limits on the mediator mass}

In this appendix, we examine how the exclusion limits on the DM-proton cross section vary with the mass of the vector mediator, $m_{A'}$.

\begin{figure}[]
\centering
\includegraphics[width=8cm]{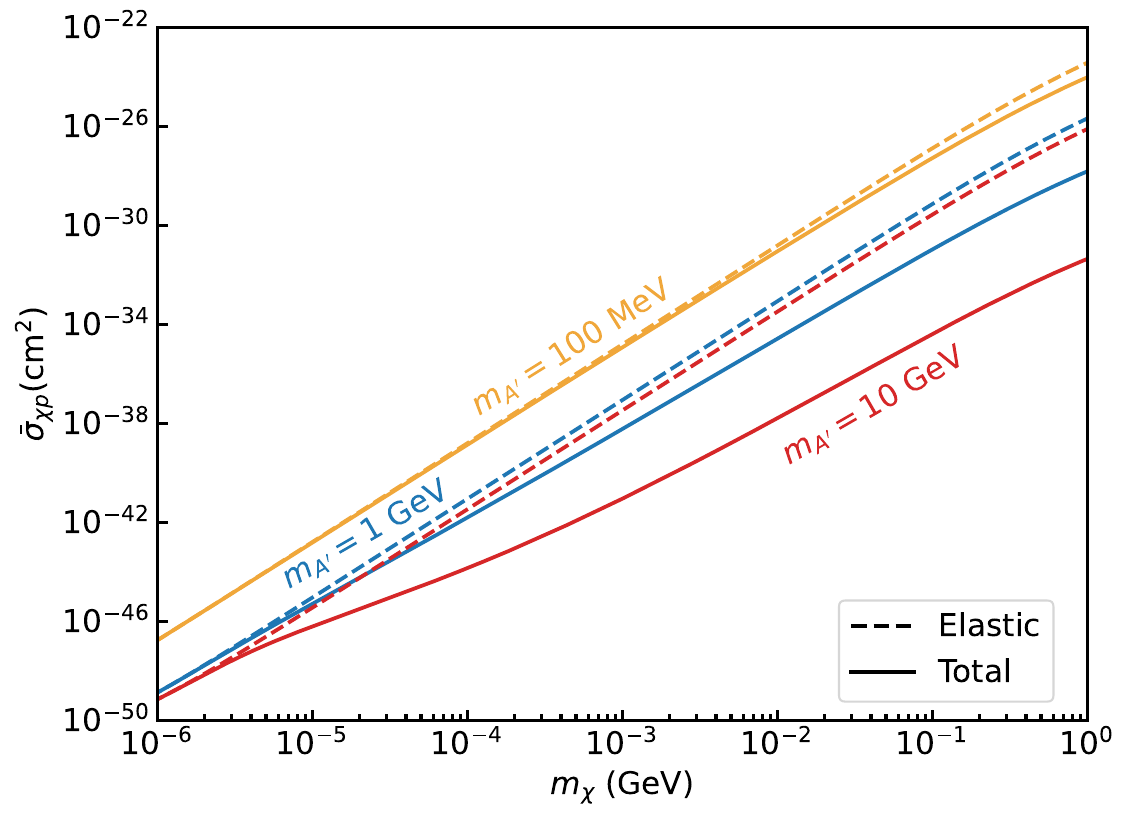}
\caption{Exclusion limits on the DM-proton cross-section as a function of DM mass for three mediator masses: $m_{A'}=100~\mathrm{MeV}$ (yellow), $1~\mathrm{GeV}$ (blue), and $10~\mathrm{GeV}$ (red), at a fixed CR energy $T=10^6~\mathrm{GeV}$. Dashed lines consider only elastic scattering (Elastic), while solid lines include both elastic and deep inelastic scattering (Total).}
\label{fig:massrange}
\end{figure}

 Fig.~\ref{fig:massrange} compares the exclusion limits for three representative mediator masses: $m_{A'} = 100~\mathrm{MeV}$, $1~\mathrm{GeV}$, and $10~\mathrm{GeV}$, at a fixed CR kinetic energy $T = 10^6~\mathrm{GeV}$. The limits are derived using the same cooling criterion adopted in the main text: $t_{\chi p} \leq C t_{\mathrm{SM}}$ with $C=0.1$. The change in the exclusion limits follows the scaling of the effective cross section: $\bar{\sigma}_{\chi p} \propto g_\chi^2 g_p^2 / m_{A'}^4$. For a fixed coupling product, a heavier mediator leads to a smaller total cross section. The contribution of DIS to the energy loss rate grows with $m_{A'}$, which results in stronger exclusion limits. This physical trend is directly reflected in our numerical results. The separation between the elastic-only and the total (elastic + DIS) limits grows with $m_{A'}$, reflecting the increasing role of DIS for heavier mediators. Moreover, in the very light DM mass region ($m_{\chi} \sim 10^{-6}$~GeV), the limits for $m_{A'}=1~\mathrm{GeV}$ and $10~\mathrm{GeV}$ converge and are significantly stronger than those for $m_{A'}=100$~MeV. This confirms that our stringent constraints are realized in the heavy mediator regime.

We therefore conclude that, while the exclusion limits presented in the main text are derived under the heavy mediator assumption, they can be rescaled for different mediator masses and couplings using the scaling relations outlined above. As discussed in this work, our claim of competitive constraints relative to other probes is more robust for heavy mediators.

\end{document}